%% file: main.tex
\documentclass[manuscript, screen, nonacm,review=false,timestamp=false]{acmart}
\AtBeginDocument{%
  \providecommand\BibTeX{{%
    \normalfont B\kern-0.5em{\scshape i\kern-0.25em b}\kern-0.8em\TeX}}}

\makeatletter
\let\@authorsaddresses\@empty
\makeatother

\usepackage{eurosym}
\usepackage{tabularx}
\usepackage{paralist}

\setcopyright{rightsretained} 
\acmDOI{10.1145/3498366.3505835}

\begin{document}

\title[SaL-Lightning Dataset]{SaL-Lightning Dataset: Search and Eye Gaze Behavior, Resource Interactions and Knowledge Gain during Web Search}
\author{Christian Otto}
\orcid{0000-0003-0226-3608}
\affiliation{%
	\institution{TIB -- Leibniz Information Centre for Science and Technology}
	\city{Hannover}
	\country{Germany}
}
\email{christian.otto@tib.eu}

\author{Markus Rokicki}
\affiliation{%
	\institution{L3S Research Center, Leibniz University Hannover}
	\city{Hannover}
	\country{Germany}
}
\email{rokicki@l3s.de}

\author{Georg Pardi}
\orcid{0000-0002-7276-4099}
\affiliation{%
	\institution{IWM -- Leibniz-Institut f\"ur Wissensmedien}
	\city{T\"ubingen}
	\country{Germany}
}
\email{g.pardi@iwm-tuebingen.de}

\author{Wolfgang Gritz}
\orcid{0000-0003-1668-3304}
\affiliation{%
	\institution{TIB -- Leibniz Information Centre for Science and Technology}
	\city{Hannover}
	\country{Germany}
}
\email{wolfgang.gritz@tib.eu}

\author{Daniel Hienert}
\orcid{0000-0002-2388-4609}
\affiliation{%
	\institution{GESIS -- Leibniz Institute for the Social Sciences}
	\city{Cologne}
	\country{Germany}
}
\email{daniel.hienert@gesis.org}

\author{Ran Yu}
\orcid{0000-0002-1619-3164}
\affiliation{%
	\institution{Data Science \& Intelligent Systems (DSIS), University of Bonn}
	\city{Bonn}
	\country{Germany}
}
\email{ran.yu@uni-bonn.de}

\author{Johannes von Hoyer}
\orcid{0000-0003-4306-0591}
\affiliation{%
	\institution{IWM -- Leibniz-Institut f\"ur Wissensmedien}
	\city{T\"ubingen}
	\country{Germany}
}
\email{j.hoyer@iwm-tuebingen.de}

\author{Anett Hoppe}
\orcid{0000-0002-1452-9509}
\additionalaffiliation{%
	\institution{L3S Research Center, Leibniz University Hannover}
	\city{Hannover}
	\country{Germany}
}
\affiliation{%
	\institution{TIB -- Leibniz Information Centre for Science and Technology}
	\city{Hannover}
	\country{Germany}
}
\email{anett.hoppe@tib.eu}

\author{Stefan Dietze}
\affiliation{%
	\institution{GESIS -- Leibniz Institute for the Social Sciences}
	\city{Cologne}
	\country{Germany}
}
\additionalaffiliation{%
	\institution{Heinrich-Heine-University Düsseldorf}
	\city{Düsseldorf}
	\country{Germany}
}
\email{stefan.dietze@gesis.org}

\author{Peter Holtz}
\orcid{0000-0001-7539-6992}
\affiliation{%
	\institution{IWM -- Leibniz-Institut f\"ur Wissensmedien}
	\city{T\"ubingen}
	\country{Germany}
}
\email{p.holtz@iwm-tuebingen.de}

\author{Yvonne Kammerer}
\orcid{0000-0003-4341-517X}
\affiliation{%
	\institution{IWM -- Leibniz-Institut f\"ur Wissensmedien}
	\city{T\"ubingen}
	\country{Germany}
}
\email{y.kammerer@iwm-tuebingen.de}

\author{Ralph Ewerth}
\orcid{0000-0003-0918-6297}
\additionalaffiliation{%
	\institution{L3S Research Center, Leibniz University Hannover}
	\city{Hannover}
	\country{Germany}
}
\affiliation{%
	\institution{TIB -- Leibniz Information Centre for Science and Technology}
	\city{Hannover}
	\country{Germany}
}
\email{ralph.ewerth@tib.eu}
\renewcommand{\shortauthors}{Otto et al.}

\begin{abstract}
The emerging research field Search as Learning investigates how the Web facilitates learning through modern information retrieval systems.
SAL research requires significant amounts of data that capture both search behavior of users and their acquired knowledge in order to obtain conclusive insights or train supervised machine learning models. However, the creation of such datasets is costly and requires interdisciplinary efforts in order to design studies and capture a wide range of features. In this paper, we address this issue and introduce an extensive dataset based on a user study, in which $114$ participants were asked to learn about the formation of lightning and thunder. Participants' knowledge states were measured before and after Web search through multiple-choice questionnaires and essay-based free recall tasks. To enable future research in SAL-related tasks we recorded a plethora of features and person-related attributes. Besides the screen recordings, visited Web pages, and detailed browsing histories, a large number of behavioral features and resource features were monitored. We underline the usefulness of the dataset by describing three, already published, use cases.
\end{abstract}



\maketitle

\input{1_introduction}
\input{3_dataset_description}
\input{4_usecases}
\input{5_conclusion}

\begin{acks}
This work is financially supported by the Leibniz Association, Germany (Leibniz Competition 2018, funding line "Collaborative Excellence", project SALIENT [K68/2017]).
\end{acks}

\clearpage
\newpage
\bibliographystyle{ACM-Reference-Format}
\bibliography{references}

\end{document}

%% file: 1_introduction.tex
\section{Introduction}
\label{sec:introduction}
\textit{Search as learning} (SAL) is an interdisciplinary field that combines the insights of (multimedia) information retrieval, human-computer interaction, and cognitive psychology. 
Its objective is to understand and support individual learning processes during Web search. 
Research questions concern 
\begin{inparaenum}[(1)]
 \item detection and prediction of learning processes taking place,
 \item assessment of resource suitability for learning in general and for specific users,
 \item adapted retrieval and ranking for optimized learning efficiency, and
 \item classification of users and user knowledge from in-session behavior.
\end{inparaenum}

Research on this kind of questions relies on study-based data that has to capture (a) search behavior of various nature and (b) knowledge metrics of users (through pre- and postests). As they have to be conducted in controlled environments, their design and execution is costly. They have to reflect realistic scenarios in order to be indicative for real-life applications: A restriction of web pages or intrusive recording equipment 
Second, assembling meaningful questionnaires for pre- and post-tests, for instance, is a challenging task, as topic domain and item difficulty have to be well calibrated.
Lastly, the logging process itself is non-trivial since available software, to our knowledge, usually only covers part of the features of interest. Therefore handcrafted, non-intrusive logging mechanisms need to be implemented manually.

To the best of our knowledge, there are currently only two SAL-focused datasets available: 
Proa\~no-R\'{i}os and Gonz\'{a}lez-Ib\'{a}\~nez\cite{proano2020} provide a set of 83 expert-generated learning paths on a diversity of topics. Each expert assembles a set of three web resources useful towards a certain learning goal, including a justification of their choice.
However, data on real-world user behavior is not included.
Gadijaru et al.~\cite{gadiraju2018chiir} present a dataset comprised of 420 crowdsourced learning sessions, investigating the information needs on the search behavior and knowledge gain of users. Our dataset improves their contribution by presenting data gathered in a controlled lab study and it captures behavioral, resource, and gaze data.
Other resources focus on either search or learning: 
(1) \textit{Search Focus} -- Datasets for the conception and optimization of search systems provide the basis for improved automatic analysis of queries~\cite{hagen2017,hagen2010}, identification of user tasks~\cite{volske2019, hagen2013}, the influence of found resources on the user's viewpoints~\cite{alkhatib2020}, and novel interaction methods such as conversational search~\cite{penha2019};
(2) \textit{Learning focus} -- Datasets from the educational domain often explore recommendation tasks \cite{verbert2011}, provide data on user behavior in restricted learning environments~\cite{grolimund2017} or specific instructional practices~
\cite{warschauer2021}. Finally, there is an active area of research on predicting the memorability of visual resources~\cite{isola2011,cohendet2019} and the impact of resource modality on learning success~\cite{hlousek2020}.
The scope of these datasets is usually limited to a single feature type; none of them collects user behavior information in a realistic and open, learning-related Web search scenario. 

In this paper, we contribute to SAL and related areas by providing a unique dataset to overcome some of the aforementioned limitations. Our SaL-Lightning dataset contains detailed recordings, pre- and post-knowledge assessments of $114$ participants, interaction data on real-world search behavior, as well as resource features of a user study. This data diversity has the potential to help researchers answer diverse questions tied to the entire online learning framework, from individual psychological aspects, over usability tests and data visualization over retrieval and ranking issues existing in the technology that enables this process. We prove this claim by presenting three already published works.

The remainder of the paper is structured as follows. 
First, we go into great detail regarding the data acquisition process before laying out how we curated the gathered information (Section \ref{sec:dataset}). 

Afterward, we showcase already existing use cases (Section \ref{sec:usecases_limitations}). Section \ref{sec:conclusion} concludes with a summary and possible future work. 

%% file: 3_dataset_description.tex
\section{Dataset Description}
\label{sec:dataset}

This section gives an in-depth explanation of the dataset acquisition process:
\begin{inparaenum}[(1)]
\item the initial user study in Section~\ref{sec:userstudy}, 
\item the technical environment in Section~\ref{sec:technical_env} and 
\item the detailed description of the different data subsets in Section~\ref{sec:dataset_structure}.
\end{inparaenum}

\subsection{User Study}
\label{sec:userstudy}
\subsubsection{Participants and Task}
The participants (N=114), German speaking university students (95 female, $\mu_{age}$=$22.88$, $\sigma_{age}$=$2.93$) from different majors were asked to solve a realistic learning task to understand the principles of thunderstorms and lightning. This topic has been used before to study multimedia learning (e.g., ~\cite{mayer1998split,schmidt2011role}) and has been chosen since it requires learners to gain knowledge about different physical and meteorological concepts and their interplay, i.e., they need to learn about causal chains of events and acquire declarative as well as procedural knowledge \cite{anderson2001taxonomy}.
The acquisition of information about such task a can be accomplished through studying different representation formats, such as text, pictures, videos, or combinations of those. This circumstance is beneficial for our goal to get a general idea about optimal multimedia learning resource design, especially in SAL scenarios.


%
\subsubsection{Procedure and Measurements}
The experiment consisted of an online and a laboratory part. In the online part, which had to be completed around one week before the lab appointment, participants had to respond for the first time to the 10-item multiple-choice and 4-item transfer knowledge test based on previous work~\cite{schmidt2011role}. Further, participants worked on questionnaires regarding their achievement motivation \cite{engeser2005messung} and their Web-specific epistemic justification beliefs \cite{braaten2019validation}. At the lab appointment, participants first completed tests assessing their reading comprehension \cite{schneider2007lgvt} and working memory capacity \cite{conway2005working}. 
The participants were asked to write a first essay (t1) about the topic of the formation of thunderstorms and lightning. 
Afterward, they were instructed to learn about this specific topic by searching the Web in a self-regulated manner. Participants were informed about the time limit of max. 30 minutes for their web search, and 
that they could also end the task early. They were encouraged to use every kind of Web resource they would like. After the learning phase, they were asked to write again everything what they now knew about the topic in a free essay (t2) format. Lastly, the participants were asked to answer the multiple-choice questionnaires (t2) again followed by a questionnaire assessing task engagement \cite{matthews2002fundamental} and cognitive reflection tasks \cite{frederick2005cognitive}.

\subsection{Technical environment}
\label{sec:technical_env}

All search and learning activities of participants were conducted within a tracking framework consisting of two layers. The first layer was the SMI (SensoMotoric Instruments) ExperimentCenter (3.7) software environment that enabled us to track participants' eye movements as well as their activities during Web search in the formats of screen recordings and navigation log files. 

For the second layer we utilized browser plugins to gather resources of all visited HTML files and adapted Talibi et al's method \cite{csedu18} to track navigation and interaction data (e.g., mouse movements). For more details we refer to~\cite{otto2021predicting}.

\subsection{Dataset Structure}
\label{sec:dataset_structure}

The following section describes the information per user provided by the individual data subsets. Apart from the screen recordings and HTML data, which we cannot make publicly available due to licensing restriction, all dataset parts are available under \url{https://doi.org/10.25835/0062363}.

\subsubsection{Resource Data - Screen Recordings} 
\label{sec:screen_recordings}
The screen recordings show the entire search process of the participants over the duration of the study and are aligned with the provided logs (Section~\ref{sec:timeline}) and HTML data (Section~\ref{sec:html_data}). The screen recording's video format is MP4, and they have been recorded with a resolution of $1280x720$ at $30$ frames per second. The audio track is not included. They are not longer than $30$ minutes and start at the point in time the learning session starts. We manually cut the start of the video that showed the participants entering their session IDs. 

\subsubsection{Resource Data - HTML} 
\label{sec:html_data}
Since online content is not persistent, to achieve our goal to enable research on the actual data seen by the participants, we decided to record the content of each visited website, including but not limited to *html, *css, *js, and image files. Due to technical difficulties this process was not entirely successful, forcing us to fill in the gaps at later points in time. In detail, we managed to capture $87.9\%$ of the data at the time of the study, another $4\%$ in March 2020 and finally, another $2.5\%$ in September 2021. For the remaining $5.6\%$ (181 URLs) we were not able to record any data. With very few exceptions (a few websites that are not available anymore) these were search engine result pages from Google and YouTube that do not contain any learning relevant information, and when crawled at a later point in time, differ strongly from the original. For these two reasons we decided to exclude them from the dataset. For full transparency we disclosed the date of acquisition in the provided timeline.

\subsubsection{Behavioral Data - Browsing Timeline} 
\label{sec:timeline}

Each participant's browsing log is represented by one TSV (tab separated value) file as outlined by Table~\ref{tab:timeline}, chronologically displaying the visited websites with a timestamp in seconds passed since the start of the session.
Additionally, we disclose the path to the respective HTML files and their acquisition date as mentioned in Section~\ref{sec:html_data}.

\begin{table}[!ht]
    \centering
    \begin{tabular}{|c|c|c|c|c|}
        \hline
        p\_id & timestamp & url & html\_files & date\_of\_acquisition \\
        \hline
    \end{tabular}
    \caption{The fields (columns) in the timeline file associating each displayed web resource with a directory of HTML files and its date of acquisition.}
    \label{tab:timeline}
\vspace{-0.5cm}
\end{table}

\subsubsection{Behavioral Data - Gaze} 
\label{sec:gaze_data}

As mentioned in Section~\ref{sec:technical_env}, we used an eye-tracker to record the learner's eye movements over the course of the Web search session. 
Similar to the sections before, this subset contains one TSV file for each participant, chronologically listing the coordinates of the left and right pupil with millisecond precision. Additionally, the URL visible at that point in time is displayed (Table~\ref{tab:gaze_columns}).

\begin{table}[!ht]
    \centering
    \begin{tabular}{|c|c|c|c|c|c|c|}
        \hline
        p\_id & timestamp & left\_x & left\_y & right\_x & right\_y & url \\
        \hline
    \end{tabular}
    \caption{The fields (columns) in the gaze data files, chronologically displaying the gaze coordinates for each eye.}
    \label{tab:gaze_columns}
\vspace{-0.5cm}
\end{table}

\subsubsection{Behavioral Data - Browsing Events}
\label{sec:events}

The investigation of the Web search behavior requires detailed logs about the learners' interactions with a website, going beyond logging what type of resources they visited. We recorded over 1 million user interaction events of 11 different types. 
The \textit{focus}, \textit{blur}, and \textit{beforeunload} describe whether a website has come into focus, lost focus, or is about to be closed. 
The \textit{resize} event tracks if a participant chose to resize the current browser window and captures the resulting window size in the value column encoded as pixel sizes $x|y$. Similarly, if a learner scrolled on a website, a \textit{scroll} event is triggered and we log the scroll distance in vertical and horizontal direction in the value column as $vertical|horizontal$. The \textit{mousemove} event tracks the learners' mouse movements by logging x and y coordinates in the respective columns. Mouse clicks were captured in the \textit{click} event, tracking their location (x and y columns) and the clicked HTML element in the target column as XPath. We recorded keypresses in the \textit{keypressed} event, but omitted recording the key values due to privacy reasons in case the learner chose to login somewhere during the session. However, the chosen queries are available in the URLs of the respective search engines.
Lastly, we captured \textit{copy} and \textit{paste} events and, if available, the intended target elements. The structure of the data record per event is displayed in Table \ref{tab:event_columns_events}.

\setlength\tabcolsep{4pt}
\begin{table}[!ht]
    \centering
    \begin{tabular}{|c|c|c|c|c|c|c|c|c|}
        \hline
        p\_id & timestamp & track\_id & type & value & x & y & target & url \\
        \hline
    \end{tabular}
    \caption{The columns in the event data files for each participant chronologically displaying the browsing interaction events.}
    \label{tab:event_columns_events}
\vspace{-0.5cm}
\end{table}

\subsubsection{Behavioral Data - Browsing Tracks}
\label{sec:tracks}
The browser tracking tool associates events to websites by means of \textit{tracks}. Upon navigating to a website, a track is created and exists until the user navigates somewhere else within the same browser tab or closes it. This setup is geared towards realistic search sessions with multiple concurrent tabs. For each track, our dataset contains the time of creating the track, URL, and title of the website, as well as the viewport dimensions. Additionally, the data contains the lifetime of the track, as well as the time the track was active, i.e., it was displayed to the user in the active browser tab.  

\begin{table}[!ht]
    \centering
    \begin{tabular}{|c|c|c|c|c|}
        \hline
        p\_id & timestamp & track\_id & url & title \\
        \hline
    \end{tabular}
    \begin{tabular}{|c|c|c|c|}
        \hline
        viewport width & viewport height & time stay & time active \\
        \hline
    \end{tabular}
    \caption{The columns in the track data files, capturing information such as URL and active time for a visited website.}
    \label{tab:browsing_tracks}
\vspace{-0.5cm}
\end{table}

\subsubsection{Knowledge Data and Questionnaires}
\label{sec:personal_data}
As mentioned in 2.1.2, we measured the knowledge state of learners at multiple points in time. Additionally, through questionnaires and tests, we captured cognitive abilities and assessments of participants across the study. Thereby, several sub datasets were generated for which we provide the documentations with explanations of measured variables and, if possible, the original German items. This section will give brief explanations of these files, while detailed documentation can be found in the dataset. 

\textbf{demo\_knowledge\_sum.csv:} This file contains demographic information of participants and the summary of the knowledge-related scores (multiple-choice, essay) and cognitive abilities (working memory capacity, reading comprehension, cognitive reflection) (Table~\ref{tab:knowledge_table}). Reading comprehension was measured through a standardized German screening instrument for adolescents and young adults. Working memory capacity was measured through a reading span task. Detailed information is provided by Pardi et al.~\cite{pardi2020role}.

\begin{table}[!ht]
\centering
\begin{tabularx}{0.75\linewidth}{|l|X|}
\hline
 \textbf{Feature} & \textbf{Description} \\ \hline\hline
p\_id                       & Participant ID \\\hline
d\_sex                      & 1= female; 2 male\\\hline
d\_age                      & Age of participant\\\hline
d\_field                    & Field of study\\\hline
d\_no\_sem                  & Number of semesters\\\hline
d\_lang                     & Mother tongue\\\hline
k\_mc\_sum\_t1              & \# of correct mc questions before search\\\hline
k\_mc\_sum\_t2              & \# of correct mc questions after search\\\hline
kg\_mc                      & Knowledge gain multiple choice questions\\\hline
essay\_C1                   & \# of correct concepts before search\\\hline
essay\_C2                   & \# of correct concepts after search\\\hline
kg\_essay                   & Knowledge gain essays\\\hline
LGVT\_speed                 & \# of words read\\\hline
LGVT\_core                  & Points for correctly solved sentences\\\hline
WMC\_Recalls                & \# of correctly recalled sets\\\hline
WMC\_Sentence               & \# of correctly solved sentences\\\hline
CRT\_sum                    & \# of correctly solved cognitive reflection tasks\\\hline
\end{tabularx}
\caption{The columns in the demo\_knowledge\_sum data files. One participant per row.}
\label{tab:knowledge_table}
\end{table}

\noindent\textbf{mc\_data.csv:} This file contains the scores for all multiple-choice questions answered by participants before the lab session (t1) and after the search in the lab (t2). Includes also the confidence rating of participants for each question and the information if the answer was guessed. 

\noindent\textbf{essay\_data.csv:} This file contains the raw essays written by the participants before (t1) and after the search (t2).

\noindent\textbf{internet\_specific\_epistemic\_justification.csv:} This file contains the measured Web-specific epistemic justification based on a translated version of \cite{braaten2019validation}. 

\noindent\textbf{selfassessment\_data.csv:} To measure participant´s self-assessed performance on the knowledge tests, we used both global self-assessment (estimated numbers of items answered correctly, estimated placement as compared to others and perceived ability to explain the concepts of the learning topic) as well as local on-item confidence rating, indicating how confident participants were that their given answer was correct.

\noindent\textbf{CRT\_data.csv:} To measure an individual´s tendency for cognitive reflection, participants worked on five items of the cognitive reflection task (CRT~\cite{frederick2005cognitive}) translated into German. Within the dual-process model of reasoning, there is a distinction between faster responses with little deliberation and slower and more reflective responses. Solving more of the CRT items shows a higher disposition for the latter one, i.e., reflective cognition.  

\noindent\textbf{achievement\_data.csv:} We used the German version of the achievement motives scale \cite{engeser2005messung} to measure hope of success (HS) and fear of failure (FF). This scale contains 10 items that are rated on a scale from 1 to 4, assessing those two dimensions. The achievement motive of an individual describes the general tendency to approach or avoid success in an evaluative situation. The HS score is associated with a range of variables beneficial for learning success, such as, performance in reasoning, persistence, or  task enjoyment.

\noindent\textbf{dssq\_data.csv:} The Dundee Stress State Questionnaire~\cite{matthews2002fundamental} measures subjective states in a performance context. Participants had to indicate their agreement to 7 items with labeled endpoints immediately after the learning phase. The mean score of those items indicates an individual´s self-reported task engagement during the learning phase. Differences in task engagement can act as a moderator for task performance.

%% file: 4_usecases.tex
\section{Use Cases}
\label{sec:usecases_limitations}
The following paragraphs describe three studies that were conducted based on the SaL-Lightning dataset.

\subsubsection*{Knowledge Prediction}
\label{sec:knowledge_prediction}

One of the most sought after tasks in SAL is the prediction of the learning outcome given a set of learning resources and, optimally, considering the learner's individual needs. The prediction of the user's knowledge would allow retrieval algorithms to present more accurate search results given a set of resources for a respective topic. 
Otto et al.~\cite{otto2021predicting} investigated how the layout and content of the multimedia data on the websites influence, besides behavioral and resource features, the knowledge gain of the participants.

Further, they conducted correlation analyses of (multimodal) resource features and pre-/post-knowledge scores, which might indicate whether certain resources or resource types (e.g., videos) are  more frequently used by novices, intermediates, or experts. 
Since our pre- and post-study questionnaires have measured the knowledge state both using multiple-choice questions and essays, the dataset can be used to advance the state of the art in this research area.

\subsubsection*{Cognitive Abilities and Search Tasks}
\label{sec:cognitive_judgments}

Pardi et al.~\cite{pardi2020role} investigated the connection between behavioral data, namely the time spent on different information resources, and the learners' cognitive abilities and learning outcome. Therefore, they classified based on the URL and screen recordings the time participants spent on text-dominated websites (potentially accompanied by images) and the time spent on online videos. Furthermore, they analyzed the assessed working memory capacity and reading comprehension of participants in connection to their learning outcome. The learning outcome was derived from the pre- and post essays users wrote before and after their web search. Based on these data, the correlations between learning outcome (pre, post), cognitive abilities (working memory, reading comprehension), and time spent on resources (search activities, on video, on text-dominated websites) were analyzed.

\subsubsection*{Metacognitive Judgments in SAL}
\label{sec:Metacognitive Judgments in SAL}

Hoyer et al.~\cite{Hoyermeta2019} used the global and local confidence ratings to investigate how the evaluation of one´s own knowledge changes after a learning phase. The authors found that after the 30-minutes self-regulated online learning phase, learners are able to gain knowledge and are in general capable of accurately estimating their knowledge on measures of global confidence ratings. However, there was also an increase in overclaiming, which indicates that more knowledge is claimed after learning. Additionally, the authors report an unexpected false certainty effect indicated by increased local confidence  ratings given to incorrectly answered questions after the learning phase. Since accurate local metacognitive judgments are essential for controlling learning processes, this result points to a possible detrimental effect of short online learning.

%% file: 5_conclusion.tex
\section{Conclusions and Future Work}
\label{sec:conclusion}

In this paper, we have presented a comprehensive, multilateral dataset for research in the field of Search as Learning. It enables interdisciplinary research in various disciplines related, but not limited to, multimodal information retrieval and learning psychology, and usability. The potential is underlined by three already published publications that made use of the SaL-Lightning dataset. 
Our plan for the future is to explore further interdisciplinary topics related to Search as Learning, in addition to the ones already mentioned. Besides the exploration of psychological phenomena based on the provided knowledge metrics, a variety of computer science applications are also unexplored, for instance, multimedia resource recommendation, web document layout analysis, analysis of user behavior during search, and query refinement.